\documentclass[a4paper, 12pt]{article}

\newtheorem{theorem}{Theorem}

\newtheorem{definition}{Definition}
\begin{document}

\title{Polarized Electrogowdy spacetimes censored}

\author{Ernesto Nungesser \\ 
Max-Planck-Institut f\"ur Gravitationsphysik\\
Albert-Einstein-Institut\\
Am M\"uhlenberg 1\\
14476 Potsdam, Germany}

\maketitle
\date{}

\begin{abstract}
A sketch of the proof of strong cosmic censorship is presented for a class of solutions of the Einstein-Maxwell equations, those with polarized Gowdy symmetry. A key element of the argument is the observation that by means of a suitable choice of variables the central equations in this problem can be written in a form where they are identical to the central equations for general (i.e. non-polarized) vacuum Gowdy spacetimes. Using this it is seen that the results of Ringstr\"{o}m on strong cosmic censorship in the vacuum case have implications for the Einstein-Maxwell case. Working out
the geometrical meaning of these analytical results leads to the main
conclusion.
\end{abstract}

\section{Strong Cosmic Censorship (SCC)}

From the singularity theorems of Hawking and Penrose we know that singularities appear under general physically reasonable assumptions, but we do not know what happens in the singularities without a theory of quantum gravity.
So if the singularities can influence what is outside, there would be a breakdown of predictability. A conjecture to solve this problem is the strong cosmic censorship hypothesis of Penrose \cite{P2}. We will use the following definition of SCC taken from \cite{alan}:
\begin{definition}(SCC)
The maximal Cauchy development of all initial data for the Einstein-matter system belonging to an open dense subset is inextendible provided the matter model is well-behaved.
\end{definition}
This ensures that solutions which are non-generic or ``unstable'' and with a ``badly-behaved'' matter model are not taken into account (see chapter 9.4 of \cite{alan} for a discussion of that). For the reader interested in non-generic solutions within the Gowdy-class we recommend \cite{Q}. Since the maximal Cauchy development is the largest region of spacetime which is uniquely determined by initial data, if this region is inextendible, general relativity would remain a deterministic theory. In practice one has to look at the structure of the singularities (initial sinuglarity) or to show that there do not exist (in the future for example). Strictly speaking not related to SCC in general, but assuming symmetries one can look what can be shown and often the techniques used can be generalized at least in principle.

\section{Gowdy symmetry}

Gowdy spacetimes are vacuum spacetimes which model closed universes filled with gravitational waves of two polarizations. These spacetimes are a good toy model for the understanding of inhomogeneous and anisotropic cosmological models. Since we will assume some matter content, in particular a Maxwell field, we will use the terminology of Gowdy-symmetric spacetimes instead of Gowdy spacetimes where the Gowdy symmetry is defined as follows:

  \begin{definition}(Gowdy symmetry)
\begin{enumerate}
  \item It is a $T^2$ $(U(1)\times U(1))$ symmetry with the group action generated by two commuting spacelike Killing 
vectors ($\frac{\partial}{\partial x}$ and $\frac{\partial}{\partial y}$)
  \item The transformation which simultaneously maps $x$ to $-x$ and $y$ to $-y$ is an
isometry
\end{enumerate}
\end{definition}
With this group action the possible spatial topologies are basically $S^3$, $S^2 \times S^1$ and $T^3$ and in the following the spatial topology is assumed to be the three-dimensional torus. A metric with Gowdy symmetry is said to be \textit{polarized} if the \textit{individual} transformations mapping $x$ to $-x$ and $y$ to $-y$ are symmetries, which has the physical interpretation that the gravitational waves have only one polarization. We will assume that our spacetime has a polarized Gowdy-symmetry.

\subsection{Polarized Gowdy metric}
It has been shown that the area spanned by the two Killing fields can be used as a time coordinate and moreover that there exists a Cauchy hypersurface of constant t. Using this \textit{areal time} coordinate the polarized Gowdy metric can be put in the following manner:
  \begin{equation}
t^{-\frac12}e^{\frac{\lambda(t,\theta)}{2}}(-dt^2+d\theta^2)+t(e^{P(t,\theta)}dx^2+e^{-P(t,\theta)}dy^2)
\end{equation}
Here $\theta$, $x$ and $y$ are periodic coordinates on $T^3$ and $\lambda$ and $P$ are smooth functions.

\section{Einstein-Maxwell equations}
\subsection{Maxwell field and new variables}
The matter content will be described by a Maxwell field which is defined by a four-potential $A_\alpha$. We choose a field which is consistent with the symmetries and denote the remaining components of the potential by 
\begin{eqnarray}
 A_2&=&\omega \\
A_3&=&\chi.
\end{eqnarray}
Before coming to the equations we introduce new variables 
\begin{eqnarray}
 \bar P&=&\frac12 (P-\log t)\\ 
\bar\lambda&=&\frac14(\lambda-\log t)-\bar P.
\end{eqnarray}
What we obtain is the same PDE-system as in the non-polarized Gowdy case! That this choice of variables linking the two cases exists was already known long time ago (see for instance \cite{CA}). This enabled us to use the work of Ringstr\"om and others obtained for the Gowdy spacetimes which has culminated in \cite{R0} and \cite{R1}.

\subsection{Basic equations for Einstein-Maxwell assuming polarized Gowdy symmetry}
The basic equations are the following:
\begin{eqnarray}
&&-\bar P_{tt}-t^{-1}\bar P_t+\bar P_{\theta\theta}=
e^{2\bar P}(-\chi_t^2+\chi_\theta^2)\label{gowdy7}                          \\
&&-\chi_{tt}-t^{-1}\chi_t+\chi_{\theta\theta}
=-2(-\bar P_t\chi_t+\bar P_\theta\chi_\theta)\label{gowdy8}                 \\
&&\bar\lambda_t=t[\bar P_t^2+\bar P_\theta^2+e^{2\bar P}
(\chi_t^2+\chi_\theta^2)]\label{gowdy9}                                     \\
&&\bar\lambda_\theta=2t[\bar P_t\bar P_\theta+e^{2\bar P}\chi_t\chi_\theta]
\end{eqnarray}
where the first two are the evolution and the last two the constraint equations. There is also a consistency condition which lead us to assume that $\omega=0$ and other equations involving higher derivatives of $\lambda$ which are not important here. Also an integral constraint coming from the fact that $\lambda$ is periodic has to be satisfied. In terms of our new variables the metric is
\begin{equation}\label{gowdymetric2}
e^{2(\bar\lambda+\bar P)}(-dt^2+d\theta^2)+t^2e^{2\bar P} dx^2+e^{-2\bar P}dy^2.
\end{equation}

\section{Central results}

\subsection{Key steps}
There is a large class of Gowdy solutions called the low velocity solutions which, when translated to the case of interest here, admit asymptotic expansions of the following form 
\begin{eqnarray}
&&\bar P(t,\theta)=-v_\infty(\theta)\log t+\ldots\label{lowvelocity1}
  \\
&&\chi(t,\theta)=q(\theta)+\psi (\theta)t^{2v_\infty}+\ldots\label{lowvelocity2}
\end{eqnarray}
where the terms omitted are lower order as $t\to 0$. $q$ and $\psi$ are smooth functions and $v_\infty(\theta)$ is called the asymptotic velocity. There are also more complicated solutions with so called true and false spikes which had to be treated separately. In order to show that the spacetime is not extendible we looked at invariants and hoped that they become unbounded. We looked at $F^{\alpha\beta}F_{\alpha\beta}$, but this did not work in general with the available information. $F^{\alpha\beta}{}^*F_{\alpha\beta}$ vanishes in our case so this invariant is not helpful. Finally it worked with the Kretschmann scalar $R_{\alpha\beta\gamma\delta}R^{\alpha\beta\gamma\delta}$. \\ The argument used to prove geodesic completeness in the vacuum case basically also applies to the Einstein-Maxwell case, proving geodesic completeness in that case too. Moreover the solutions exhibit oscillatory behaviour at late times analogous to that found in
\cite{R0} for Gowdy spacetimes.

\subsection{Central results}
In the following theorems the topology on the set of initial data used is the 
$C^\infty$ topology. ${\cal G}_c$ is a generic set of initial data satisfying an integral constraint. Data belonging to ${\cal G}_c$ are called generic. This subset is open and dense in the set of initial data in the $C^\infty$ topology.
\begin{theorem}\label{blowup}
There is an open dense subset ${\cal G}_c$ of the set of 
smooth initial data for the Einstein-Maxwell equations with polarized Gowdy 
symmetry and constant areal time such that the Kretschmann scalar tends to 
infinity along any inextendible past-directed causal geodesic.
\end{theorem}
Another result is that strong cosmic censorship holds for polarized Electrogowdy, more precisely
\begin{theorem}For data belonging to the open dense subset ${\cal G}_c$ of 
the set of smooth initial data for the Einstein-Maxwell equations with 
polarized Gowdy symmetry and constant area radius the corresponding maximal 
Cauchy development is inextendible.
\end{theorem}

The proofs and details can be found in \cite{E1} and \cite{E2}.

\section{Conclusions and Outlook}

The analysis of the structure of the sigularities in our case, the case of gravitational waves with only one polarization and an electromagnetic field has revealed to be neither simpler nor more difficult than the vacuum case of gravitational waves with two polarizations (once the PDE-system is solved of course). Generalizations of the results to polarized Gowdy-symmetric spacetimes with a Maxwell field which does not come from a potential seem quite difficult because in this case the evoultion and constraint equations are coupled. Something similar happens in the case of a positive cosmological constant and in the case of a negative cosmological constant even the existence of a global time coordinate is problematic. However there are different results for the case of a positive cosmological constant in the context of $T^2$-symmetric spacetimes (see for instance \cite{CI}, \cite{J}). The most natural generalization of our results would be the case of general solutions of the Einstein-Maxwell equations with Gowdy symmetry, i.e. without the restriction of polarization. In this direction already some results have been obtained. An energy decay has been shown by Ringstr\"om \cite{R2} which could be an important first step. The difficulties in the general case come from the fact that in the corresponding wave map formulation the target space is the complex hyperbolic space instead of the hyperbolic space. A discussion of that and clues concerning the asymptotic behaviour near the singularity can be found in \cite{MF}.\\

\textbf{Acknowledgements}\\
The author would like to thank Alan D. Rendall for a lot of helpful discussions, especially those which take place after work in the train.

\end{document}